\documentclass[twocolumn,groupaddress,aps]{revtex4-1}

\usepackage[latin9]{inputenc} %
\usepackage{amsmath} %
\usepackage{amssymb} %
\usepackage{graphicx} %
\usepackage{bm} %
\usepackage{epsfig} %
\usepackage{mathrsfs} %
\usepackage{physics} %
\usepackage{subfig} %

\begin{document}

\title{Hysteresis in a Superfluid Atom Circuit}

\author{Tieling Song and D.L.~Zhou}

\email{zhoudl72@iphy.ac.cn}

\affiliation{Beijing National Laboratory for Condensed Matter Physics,
  and Institute of Physics, Chinese Academy of Sciences, Beijing
  100190, China}

\date{\today}

\begin{abstract}
  We study a hysteresis phenomenon in a rotating BEC with a weak link
  in a quasi-one-dimensional torus by proposing a microscopic
  theoretical model including a dissipation bath. By analyzing the
  role of dissipation and the decay rates of all the energy levels, we
  are able to give a microscopic interpretation of hysteresis recently
  observed in the experiment and confirm that the hysteresis is the
  result of the presence of metastable state. In particular, we obtain
  the hysteresis loops in a quench process just as that in the
  experiment. We also find that the shape and size of the hysteresis
  loop change drastically with the strength of the link.
\end{abstract}



\maketitle

\section{Introduction}

Atomtronics~\cite{BTSe,RAPe,LAmic} is an emerging interdisciplinary
field that focuses on ultracold atom analogs of electronic circuits
and devices. A series of theoretical demonstrations
~\cite{LAmi,ACMa,LAmico,Dagha} and impressive
experiments~\cite{ARama,CRyu} with BECs, a testing bed for
atomtronics, have established this analogy, especially for Josephson
effects~\cite{BPAn,SLev,FCat}, Bloch Oscillations~\cite{OMor}.
Nevertheless, hysteresis in a superfluid atomtronic gas, which is
considered to be essential to the realization of an atomic-gas
superconducting quantum interference device (SQUID), has not been
directly observed until recently~\cite{SEck}. In this experiment, both
hysteresis and the quantization of flow have been observed in an
atomtronic circuit formed from a ring of superfluid BEC obstructed by
a rotating weak link. Just as the essential role it plays in the
electronic circuit, the realization of hysteresis in atomtronic
circuit will greatly accelerate the development of atomtronics because
the controllability of hysteresis is crucial for the requirements of
practical devices.

Hysteresis, widely used in electronic circuits, is the phenomenon
where the state of a physical system depends upon its history. The
canonical example of hysteresis in a classical system is that in
ferromagnetism. According to the Landau theory of phase
transitions~\cite{LDLa}: the energy landscape, which changes with the
applied magnetic field, is produced by calculating the energy of the
system as a function of magnetization. Hysteresis occurs when the
energy landscape has two local minima separated by an energy barrier.
At some critical field, the barrier disappears and the system has only
a global minimum.

At a rudimentary level, hysteresis manifests itself as the competition
between experimental time scales and internal time scales, the former
are determined by the frequencies of the applied perturbation and the
latter are governed by relaxation, decay and so on~\cite{VBan}. To
observe a hysteresis in a quantum system, we require that some
eigenstates of the system are metastable, namely, we can safely
neglect the decays of those states in the experimental time scale. In
this paper, we investigate the essential role that the metastable
states play in the formation of hysteresis.

The swallowtail energy loop is a generic feature of hysteresis in an
atomtronic circuit~\cite{BWuQ,EJMu,SBah,Dmit}. However, much of the
study of swallowtails is rooted in the exact solutions to the
Gross-Pitaevskii equation~\cite{JCBr,JCBro}. Recently, the dynamics
simulation in a toroidal BEC~\cite{AIYa,AIYak} is also confined to the
mean-field approximation.

In our study we introduce a microscopic model for the dissipation, and
thus we can associate the swallowtail energy loop with the existence
of metastable states. By this microscopic model, we obtain the
relative decay rates of all many-body states, which decides whether a
metastable state exists in our system or not, and leads to a
quantitative calculation of the hysteresis in our system. Our
calculations show that the interaction tends to increase the size of
the hysteresis loop, while the strength of the link tends to decrease
it, and confirm that there exists a metastable state which results in
the hysteresis loops.

\section{Theoretical Model}

\subsection{Two-mode approximation}

We consider a quasi-one-dimensional dilute gas, containing $N$ bosonic
atoms in a thin annulus of radius $R$ and cross-sectional radius
$r_{0}\ll{R}$, which rotates at frequency $\Omega$ driven by a
rotating repulsive potential $V$~\cite{JDal}. The Hamiltonian of this system
$\mathscr{S}$ in the rotating frame with frequency $\Omega$ is given by
\begin{equation}
  H_{S} = \sum_{i} \left[\frac{L_{zi}^{2}}{2MR^{2}} +
    V(\vec{r}_{i})\right] +
  \frac{1}{2} \sum_{ij} g \delta(\vec{r}_{i}-\vec{r}_{j}) -
  \Omega \sum_{i} L_{zi},
  \label{ha}
\end{equation}
where $M$ is the atomic mass, $g$ is the strength of contact
interaction, $L_{z}=-i\hbar\partial/\partial\theta$ is the angular
momentum operator, and the potential takes the form
\begin{equation}
  V =
  \begin{cases}
     V_{0}, & |\theta|\leq\theta_{0},\\
      0, & |\theta|>\theta_{0},
  \end{cases}
  \label{V}
\end{equation}
which depletes the density in a small portion of the ring and thereby
creates a weak link.

In terms of single-particle eigenstates
$\psi_{l}(\theta)=e^{il\theta}/\sqrt{2\pi}$ of $L_{z}$ with angular
momentum $l\hbar$, the Hamiltonian can be written as
\begin{align}
  H_{S} & = \frac{1}{2} \hbar \Omega_{0} \sum_{j} {(j-\bar{\Omega})}^{2}
          a_{j}^{ \dag}a_{j} + \sum_{j \neq k}
          \frac{V_{0}\sin(j-k) \theta_{0}}{(j-k) \pi} a_{j}^{\dag}a_{k}
          \nonumber\\
        & \quad {} +
          \frac{1}{2} g \sum_{jkm}
          a_{j}^{\dag}a_{k}^{\dag}a_{k-m}a_{j+m}, \label{haa}
\end{align}
where $\Omega_{0}=\hbar/MR^{2}$, $\bar{\Omega}=\Omega/\Omega_{0}$, and
$a_{j}^{\dag}$ ($a_{j}$) is the creation (annihilation) operator of
boson with angular momentum $j\hbar$.

In the experiment~\cite{SEck}, a two-step sequence is used to observe
hysteresis in a BEC of ${}^{23}$Na atoms. The BEC is firstly
prepared in either the $n=0$ or the $n=1$ circulation state by either
not rotating the weak link or by rotating it at $\Omega_{1}$. Then the
weak link is rotated at various angular velocities $\Omega_{2}$ for a
while. $V_{0}$ is ramped to a certain value $V_{1}$ in step 1 while to
a chosen $V_{2}$ in step 2. The transitions of average angular
momentum $\langle{n}\rangle=0\rightarrow1$ and
$\langle n\rangle=1\rightarrow0$ occur at different values of
$\Omega_{2}$ and form hysteresis loops.

To simulate the hysterisis in the experiment, we consider two quench
processes of parameter $\Omega$: $\Omega$ is ramped suddenly from $0$
(or $\Omega_{1}$) to a chosen $\Omega_{2}$ and then the system is
rotated at $\Omega_{2}$ for a time period such that it arrives at its
steady state. The quantity we describe the hysterisis is the average
angular momentum
\begin{equation}
  \label{eq:1}
  \hat{n} = \frac{L_{z}}{N} = \frac{\hat{n}_{1}}{N}.
\end{equation}

To simplify the discussion, we consider just two single-particle
eigenstates: the nonrotating state $\psi_{0}(\theta)$ and the state
$\psi_{1}(\theta)$ with azimuthal angular momentum $\hbar$. In fact,
since the energy of atoms is proportional to ${(j-\bar{\Omega})}^{2}$,
as long as $\bar{\Omega}<1$, atoms in either the $\psi_{0}(\theta)$ or
the $\psi_{1}(\theta)$ state will have lower energies than those in the
other states. According to the processes we consider, it is reasonable
to assume that most atoms will stay in these two levels during the
whole process. Then the Hilbert subspace is given by $|n_{0},n_{1}\rangle$
where $n_{0}$ and $n_{1}$, denoting the numbers of bosons that occupy the
states $\psi_{0}(\theta)$ and $\psi_{1}(\theta)$, satisfy $n_{0}+n_{1}=N$. The
Hamiltonian (\ref{haa}) can be approximated as
\begin{equation}
  \bar{H}_{S} = \frac{1}{2}
  [\bar{\Omega}^{2} \hat{n}_{0} +{(1-\bar{\Omega})}^{2} \hat{n}_{1}]
  + \bar{u} \hat{A} +
  \bar{g} \hat{n}_{0} \hat{n}_{1},
  \label{hab}
\end{equation}
where $\hat{n}_{0}=a_{0}^{\dagger}a_{0}$,
$\hat{n}_{1}=a_{1}^{\dagger}a_{1}$,
$\hat{A}=a_{0}^{\dagger}a_{1}+a_{1}^{\dagger}a_{0}$,
$\bar{u}=V_{0}\sin\theta_{0}/\pi\hbar\Omega_{0}$,
$\bar{g}=g/\hbar\Omega_{0}$, and $\bar{H}_{S}=H_{S}/\hbar\Omega_{0}$.

\subsection{Model of dissipation}

In the phenomenon of hysteresis, dissipation plays an essential role.
Here we introduce a simple model to describe the dissipation of our
system:
\begin{align}
  H_{R} & =  \sum_{\mu} \hbar\omega_{\mu} b_{\mu}^{\dagger}b_{\mu}, \label{hi0}\\
  H_{I} & =  \sum_{\mu} g_{\mu} \hat{A} (b_{\mu} +
          b_{\mu}^{\dagger}),\label{hi}
\end{align}
where $H_{R}$ is the Hamiltonian of the reservoir $\mathscr{R}$, and
$H_{I}$ is the interaction between the reservoir $\mathscr{R}$ and the
system $\mathscr{S}$. It is obvious that the energy of $\mathscr{S}$
can be dissipated into $\mathscr{R}$ while the particle number of
$\mathscr{S}$ is conserved.

It is worthy to point out that the dissipation model given by
Eq.~\eqref{hi0} and Eq.~\eqref{hi} is a direct generalization of the
Caldeira-Leggett model~\cite{AJLg}. In fact, our dissipation model can
be regarded as $N$ bosonic two-level atoms interact with an ensemble
of bosons, which becomes the standard Calderira-Leggett model when
$N=1$. As argued by Caldeira and Leggett, their model is generally
applicable when the coupling between the two-level system and the
environment is sufficient weak, which is the main reason why we choose
the dissipation model given in Eq.~\eqref{hi0} and Eq.~\eqref{hi}. In
our case, for example, the main physical source of the environment may
be a small number of atoms in the other modes except the two modes we
consider.

Let $|\Psi_{a}\rangle$, $|\phi_{\mu}\rangle$ be the eigenstates of
$H_{S}$ and $H_{R}$, having eigenvalues $E_{a}$ and $E_{\mu}$
respectively. Let $\rho(t)$ be the density operator at time $t$ of the
global system $\mathscr{S}+\mathscr{R}$, and $\sigma(t)$ be the
reduced density operator at time $t$ of the system $\mathscr{S}$,
which is defined by $\sigma(t)=\Tr_{R}\rho(t)$. In the basis of
eigenstates of $H_{S}$, the master equation is written as~\cite{CCoh}:

\begin{equation}
  \frac{\rm{d}\sigma_{aa}}{\rm{d}t} =\sum_{b\neq a}
  (\sigma_{bb} \Gamma_{b\rightarrow a} - \sigma_{aa} \Gamma_{a
    \rightarrow b}),
  \label{siga}
\end{equation}
where $\sigma_{aa}$ is the population of the energy level
$|\Psi_{a}\rangle$ of $\mathscr{S}$, and $\Gamma_{a\rightarrow b}$ is
the probability per unit time for the system $\mathscr{S}$ to make a
transition from level $|\Psi_{a}\rangle$ to level $|\Psi_{b}\rangle$
as a result of its coupling with $\mathscr{R}$ for $a\neq b$, which is
given by
\begin{eqnarray}
  \Gamma_{a\rightarrow b}
  & = & \frac{2\pi}{\hbar}
        \sum_{\nu} \left|\langle \phi_{\nu}, \Psi_{b} | H_{I} | \text{Vac},
        \Psi_{a} \rangle\right|^2 \delta( E_{a}- E_{\nu} - E_{b})
        \nonumber\\
  & = & \frac{2\pi}{\hbar}
        \sum_{\nu} g_{\nu}^{2} \left|\langle  \Psi_{b} | A |
        \Psi_{a} \rangle\right|^2 \delta( E_{a}- \hbar \omega_{\nu} - E_{b})
        \nonumber\\
  & = & \frac{2\pi}{\hbar} g^{2}(\omega_{ab})
        \rho(\omega_{ab}) \left|\langle  \Psi_{b} | A |
        \Psi_{a} \rangle\right|^2 \nonumber\\
  & =  & D \left|\langle  \Psi_{b} | A |
         \Psi_{a} \rangle\right|^2, \label{Raacc}
\end{eqnarray}
where the initial state of $\mathscr{R}$ is in the vacuum state
$\rho_{R}(0)=\op{\text{Vac}}$,
$\omega_{ab}=\frac{E_{a}-E_{b}}{\hbar}$, and $\rho(\omega)$ is the
energy spectrum at $\omega$. In the last line of Eq.~\eqref{Raacc}, we
assume that $\rho(\omega)g^{2}(\omega)$ does not depend on $\omega$,
and let $D=\frac{2\pi}{\hbar} g^{2}(\omega_{ab}) \rho(\omega_{ab})$.

According to Eq.~\eqref{Raacc}, the conditions for the transition rate
$\Gamma_{a\rightarrow b}>0$ are
\begin{eqnarray}
  E_{a} & \geq & E_{b},\label{eq:2}\\
  \langle\Psi_{b}|\hat{A}|\Psi_{a}\rangle & \neq & 0.\label{eq:3}
\end{eqnarray}
It means that the dissipation from $\ket{\Psi_{a}}$ to
$\ket{\Psi_{b}}$ will occur if and only if the eigenenergy of
$\ket{\Psi_{a}}$ is above that of $\ket{\Psi_{b}}$, and the transition
amplitude $\mel{\Psi_{b}}{\hat{A}}{\Psi_{b}}$ is not zero.

If an eigenstate of $\mathscr{S}$ that is not a ground state does not
significantly dissipate to any state with lower energy in the time
period we consider, then it is called a metastable state. We will show
that the existence of metastable states is essential to the formation
of hysterisis, which is discussed in Sec.~\ref{sec:3b}. It needs
to be stressed that the decay rate for the metastable state into the
other lower levels may not equal to zero exactly, but it is far less
than those of other excited levels. In the time period we consider, we
can safely neglect the decay of the metastable state. Hence, the final
state of our system is a probability distribution of the ground state
and the metastable state, and the probability distribution depends on
the history of our system, thus leading to the hysteresis, which is
shown in Sec.~\ref{sec:3d}.

\section{Analysis of hysterisis}

In this section, we first establish a general formalism for the
analysis of hysteresis in our system. Then we use it to describe four
different cases to explore the microscopic mechanism of hysterisis in
our system by steps.

Since the hysteresis is the property of the steady state of a system,
it is necessary to determine the steady state of our system by solving
the master equation~\eqref{siga}.

Let $a\in\{0,1,\ldots,N\}$, and assume that $E_{a}>E_{b}$ if $a>b$.
Then Eq.~\eqref{siga} can be rewritten as
\begin{equation}
  \label{eq:11}
  \dv{\sigma_{aa}}{t} = - \sigma_{aa} \Gamma_{a} + \sum_{a<b}
  \sigma_{bb} \Gamma_{b\rightarrow a},
 \end{equation}
where the total decay rate of the state $\ket{\Psi_{a}}$ is
\begin{equation}
  \label{eq:12}
  \Gamma_{a} = \sum_{b<a} \Gamma_{a\rightarrow b}.
\end{equation}
If $\forall a>0$, $\Gamma_{a}>0$, then the steady state will be the
ground state $\ket{\Psi_{0}}$, and our system shows no hysteresis in
the plane of $\bar{\Omega}$-$\ev{\hat{n}}$.

According to Eq.~\eqref{eq:11}, the population $\sigma_{aa}$ in the
steady state is not zero if and only if $\Gamma_{a}=0$. Obviously, the
decay rate of the ground state $\Gamma_{0}=0$. In addition, if the
state $\ket{\Psi_{c}}$ is a metastable state in our system, then also
we have $\Gamma_{c}=0$. In the steady state derived from
Eq.~\eqref{eq:11}, the population of a state without decay is
\begin{equation}
  \label{eq:13}
  \sigma_{cc}(\infty)  = \sigma_{cc}(0) + \sum_{a>c} \sigma_{aa}(0)
  p_{a\to c},
\end{equation}
where the probability of decay from $\ket{\Psi_{a}}$ to
$\ket{\Psi_{c}}$ is
\begin{equation}
  \label{eq:14}
  p_{a\to c} = \sum_{a>b_{1}>b_{2}>\cdots>b_{k}>c}  \gamma_{a\rightarrow
    b_{1}} \gamma_{b_{1}\rightarrow b_{2}} \cdots \gamma_{b_{k}
    \rightarrow c}
\end{equation}
with $\gamma_{a\rightarrow b}=\frac{\Gamma_{a\rightarrow
    b}}{\Gamma_{a}}$. In other words, $p_{a\to c}$ is the  decay
probability from $\ket{\Psi_{a}}$ to $\ket{\Psi_{c}}$ over all
possible decay paths.

The average angular momentum in the steady state is
\begin{equation}
  \label{eq:15}
  \ev{\hat{n}}  = \sum_{c:\Gamma_{c}=0} \ev{\hat{n}}{\Psi_{c}}
                 \pqty{\sigma_{cc}(0) + \sum_{a>c} \sigma_{aa}(0)
                 p_{a\to c}}.
\end{equation}
In particular, when there does not exist metastable states in our
system, Eq.~\eqref{eq:15} becomes
\begin{equation}
  \label{eq:16}
  \ev{\hat{n}} = \ev{\hat{n}}{\Psi_{0}},
\end{equation}
which implies that $\ev{\hat{n}}$ does not depend on the initial
state, and there can not exist any hysteresis in the
$\bar{\Omega}$-$\ev{\hat{n}}$ plane. Otherwise, Eq.~\eqref{eq:15}
implies that $\ev{\hat{n}}$ depends on the initial condition, and it
usually forms a hysteresis in the $\bar{\Omega}-\ev{\hat{n}}$ plane.

\subsection{The case of  $\bar{g}=\bar{u}=0$}

In this case, the eigenstates of $\bar{H}_{S}$ are
\begin{equation}
  |\Psi_{n_{0}}\rangle = |n_{0},n_{1}\rangle
  \label{Psi1}
\end{equation}
for $n_{0}=0,1,\ldots,N$ with eigenenergies
\begin{equation}
  E_{n_{0}} = \frac{1}{2} [\bar{\Omega}^{2}n_{0} + {(1 -
  \bar{\Omega})}^{2} n_{1}].
  \label{E1}
\end{equation}

\begin{figure}[htbp]
  \centering %
  \includegraphics[width=6cm]{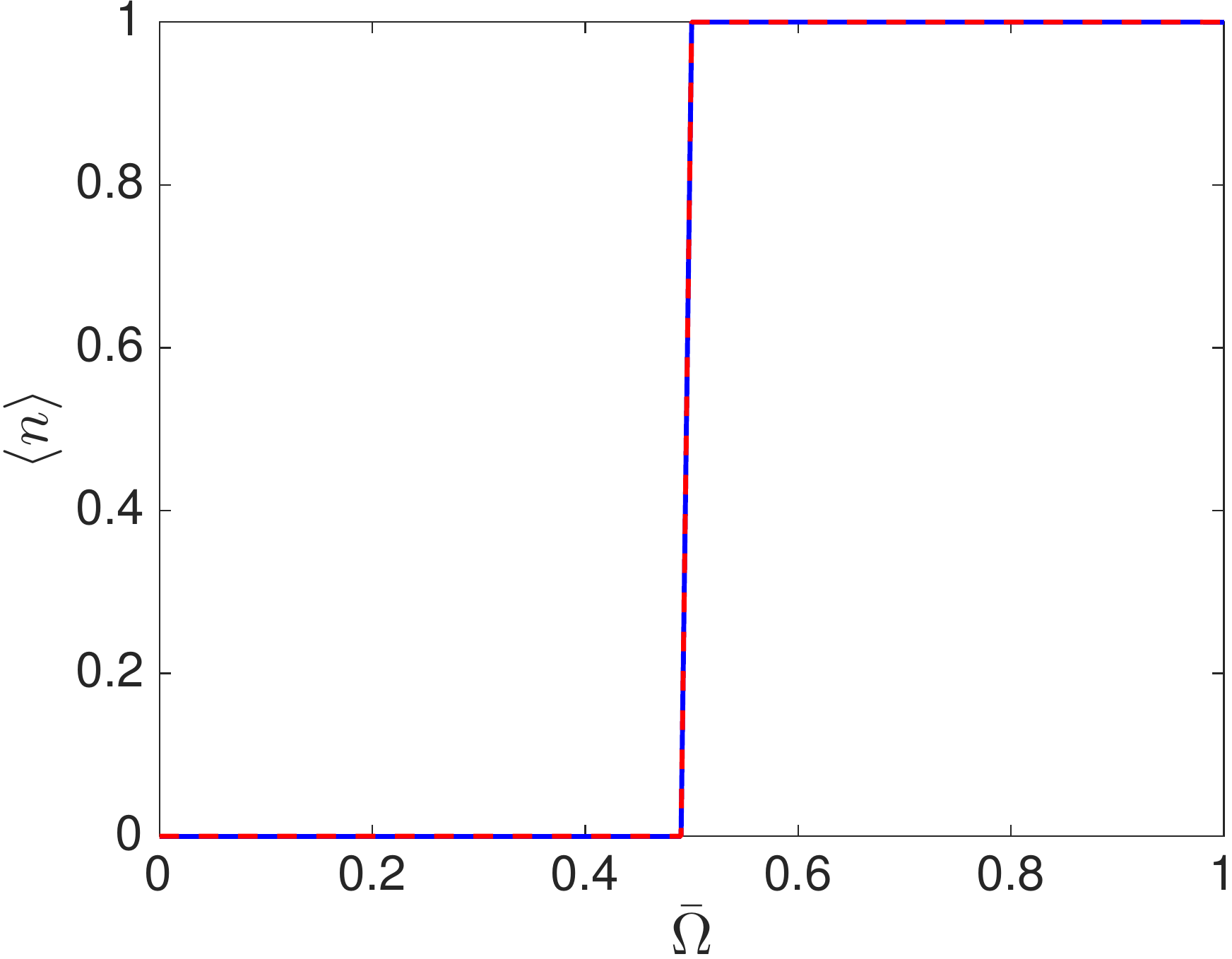}
  \caption[]{(Color online) The average momentum $\ev{n}$ via
    $\bar{\Omega}$ with parameters $N=1000$, $\bar{u}=0$, and
    $\bar{g}=0$. The solid blue line and the dashed red line
    represents different initial states $\ket{N,0}$ and $\ket{0,N}$
    respectively. }~\label{fig:g0u0}
\end{figure}

When $\bar{\Omega}<\frac{1}{2}$, $E_{n_{0}}<E_{n_{0}-1}$, and
$\ket{N,0}$ is the ground state. Because $\forall n_{0}\neq N$, the
amplitude $\mel{\Psi_{n_{0}+1}}{A}{\Psi_{n_{0}}}\neq 0$, which implies
that $\ket{\Psi_{n_{0}}}$ is not a metastable state. Hence the steady
state at $\bar{\Omega}$ ($<\frac{1}{2}$) is $\ket{N,0}$, and then the
average angular momentum $\ev{\hat{n}}=0$.

When $\bar{\Omega}>\frac{1}{2}$, $E_{n_{0}}>E_{n_{0}-1}$, and
$\ket{0,N}$ is the ground state. Because $\forall n_{0}\neq 0$, the
amplitude $\mel{\Psi_{n_{0}-1}}{A}{\Psi_{n_{0}}}\neq 0$, which implies
that $\ket{\Psi_{n_{0}}}$ is not a metastable state. Hence the steady
state at $\bar{\Omega}$ ($>\frac{1}{2}$) is $\ket{0,N}$, and then
the average angular momentum $\ev{\hat{n}}=1$. Because at any
$\bar{\Omega}$, the steady state does not depend on its history, and there
will not exist hysteresis, but the average momentum will make a
transition between $0$ and $1$ at $\bar{\Omega}=\frac{1}{2}$, see a
demonstration with $N=1000$ in Fig.~\ref{fig:g0u0}.

\subsection{The case of $\bar{g}=0,\bar{u}\protect\neq0$}

In this case, the Hamiltonian $\bar{H}_{s}$ can be diagonalized as
\begin{eqnarray}
  \bar{H}_{S}
  & = &
        \frac{1}{2}
        [\bar{\Omega}^{2} a_{0}^{\dag}a_{0} + {(1-\bar{\Omega})}^{2}
        a_{1}^{\dag}a_{1}]
        + \bar{u} (a_{0}^{\dag}a_{1} + a_{1}^{\dag}a_{0}) \notag\\
  & = &
        \begin{pmatrix}
          a_{0}^{\dag} & a_{1}^{\dag}
        \end{pmatrix}
                         \begin{pmatrix}
                           \frac{1}{2} \bar{\Omega}^{2} & \bar{u}\\
                           \bar{u} & \frac{1}{2}
                           {(1-\bar{\Omega})}^{2}
                         \end{pmatrix}
                                     \begin{pmatrix}
                                       a_{0}\\ a_{1}
                                     \end{pmatrix}
  \notag \\
  & = & \epsilon_{+} c_{+}^{\dagger} c_{+} + \epsilon_{-}
        c_{-}^{\dagger} c_{-}, \label{haa3}
\end{eqnarray}
where
\begin{equation}
  \label{eq:6}
  \epsilon_{\pm} = \frac{1}{4} \left[\bar{\Omega}^{2} + {(1 -
    \bar{\Omega})}^{2} \pm \sqrt{16\bar{u}^{2} + {(1
      - 2\bar{\Omega})}^{2}} \right],
\end{equation}
and
\begin{equation}
  \label{eq:7}
  \begin{pmatrix}
    a_{0} \\ a_{1}
  \end{pmatrix}
   =
   \begin{pmatrix}
     \cos \frac{\theta}{2} & -\sin \frac{\theta}{2} \\
     \sin \frac{\theta}{2} & \cos \frac{\theta}{2}
   \end{pmatrix}
   \begin{pmatrix}
     c_{+} \\ c_{-}
   \end{pmatrix}
\end{equation}
with $\theta = \arctan \frac{4 \bar{u}}{2 \bar{\Omega} - 1}$.

The eigenvectors and eigenvalues of $\bar{H}_{S}$ are
\begin{align}
  \label{eq:8}
  \ket{\Psi_{n_{+}}} & = \ket{n_{+},n_{-}}, \\
  E_{n_{+}} & = \epsilon_{+} n_{+} + \epsilon_{-} n_{-}.
\end{align}
Because $\epsilon_{+}>\epsilon_{-}$, we have $E_{n_{+}}>E_{n_{+}-1}$,
and $\ket{\Psi_{0}}$ is the ground state.

\begin{figure}[htbp]
  \centering %
  \includegraphics[width=6cm]{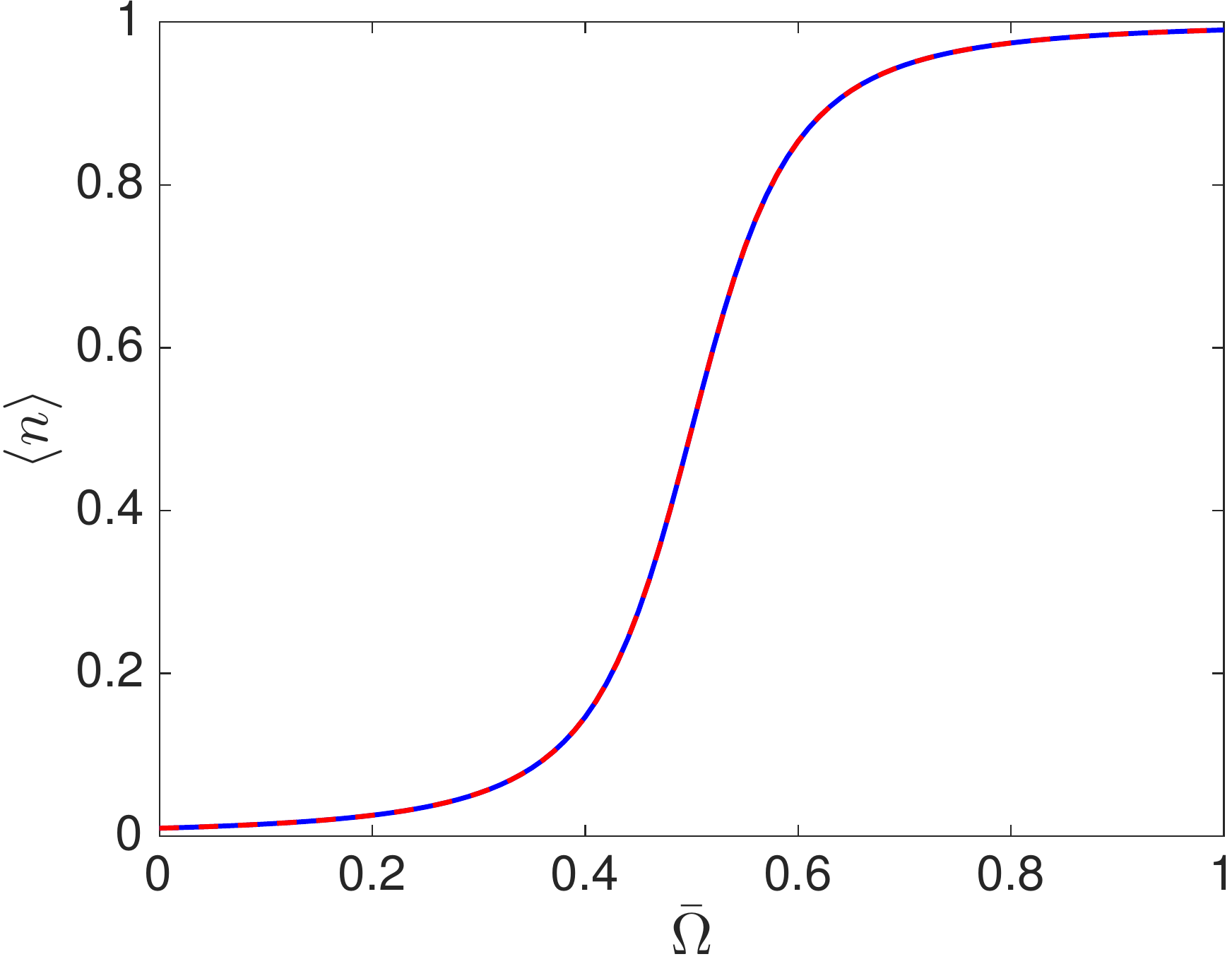}
  \caption[]{(Color online) The average momentum $\ev{n}$ via
    $\bar{\Omega}$ with parameters $N=1000$, $\bar{u}=0.05$, and
    $\bar{g}=0$.~\label{fig:g0}}
\end{figure}

According to Eq.~\eqref{eq:7}, we obtain $\forall n_{+}>0$,
\begin{equation}
  \label{eq:9}
  \mel{\Psi_{n_{+}-1}}{A}{\Psi_{n_{+}}} = \sqrt{n_{+} (N-n_{+}+1)}
  \cos \theta >0,
\end{equation}
which implies that $\ket{\Psi_{n_{+}}}$ is not metastable, i.e., there
does not exist metastable states in our system. Thus the steady state
is the ground state $\ket{\Psi_{0}}$, which does not depend on the
initial conditions. Then the average angular momentum in the ground
state is
\begin{equation}
  \label{eq:10}
  \ev{\hat{n}} = \frac{1+\cos \theta}{2} = \frac{1}{2} +
  \frac{\bar{\Omega} -1/2}{ \sqrt{16 \bar{u}^{2} + {(2 \bar{\Omega}
      -1)}^{2}}},
\end{equation}
which is numerically demonstrated in Fig.~\ref{fig:g0} with parameters
$N=1000$ and $\bar{u}=0.05$.

\subsection{The case of $\bar{g}\protect\neq0,\bar{u}=0$}
\label{sec:3b}

In this case, the eigenstates and eigenenergies of $\bar{H}_{S}$ are
\begin{eqnarray}
  |\Psi_{n_{0}}\rangle & = & |n_{0},n_{1}\rangle, \label{Psi2} \\
  E_{n_{0}} & = & \frac{1}{2} [\bar{\Omega}^{2}n_{0} +
                          {(1-\bar{\Omega})}^{2}n_{1}] + \bar{g}n_{0}n_{1}.
                          \label{E2}
\end{eqnarray}

Let us analyze the condition for the appearance of a metastable state
in our system. Because
\begin{align*}
  \mel{\Psi_{m}}{\hat{A}}{\Psi_{n_{0}}}
  & = \delta_{m,n_{0}+1} \sqrt{(n_{0}+1)(N-n_{0})} \\
  & \quad {} + \delta_{m,n_{0}-1} \sqrt{n_{0} (N-n_{0}+1)},
\end{align*}
the state $\ket{\Psi_{n_{0}}}$ is not a metastable state if and only
if $E_{n_{0}+1}<E_{n_{0}}$ or $E_{n_{0}-1}<E_{n_{0}}$. In other words,
$\ket{\Psi_{n_{0}}}$ is a metastable state if and only if $n_{0}$ is a
local minimum of the function $E_{n_{0}}$. Since
$\dv[2]{E}{n_{0}}=-\bar{g}<$0, $E_{n_{0}}$ will take a local maximum
when $\dv{E}{n_{0}}=0$. Hence only the two ends of $n_{0}$, i.e. $n_{0}=0$
and/or $n_{0}=N$, may be a local minimum of $E_{n_{0}}$. Therefore the
condition for the existence of a metastable state is $E_{0}<E_{1}$ and
$E_{N}<E_{N-1}$, which implies that
\begin{equation}
  \label{eq:4}
  \Omega_{c}^{-}<\bar{\Omega}<\Omega_{c}^{+},
\end{equation}
with
\begin{equation}
  \label{eq:5}
  \Omega_{c}^{\pm} = \frac{1}{2}\pm\bar{g}(N-1).
\end{equation}
When $\Omega_{c}^{-}<\bar{\Omega}<\frac{1}{2}$, $E_{N}<E_{0}$, and
then $\ket{N,0}$ is the ground state and $\ket{0,N}$ is the metastable
state. When $\frac{1}{2}<\bar{\Omega}<\Omega_{c}^{+}$, $E_{0}<E_{N}$,
and then $\ket{0,N}$ is the ground state and $\ket{N,0}$ is the
metastable state.

\begin{figure}[htbp]
  \centering %
  \includegraphics[width=6cm]{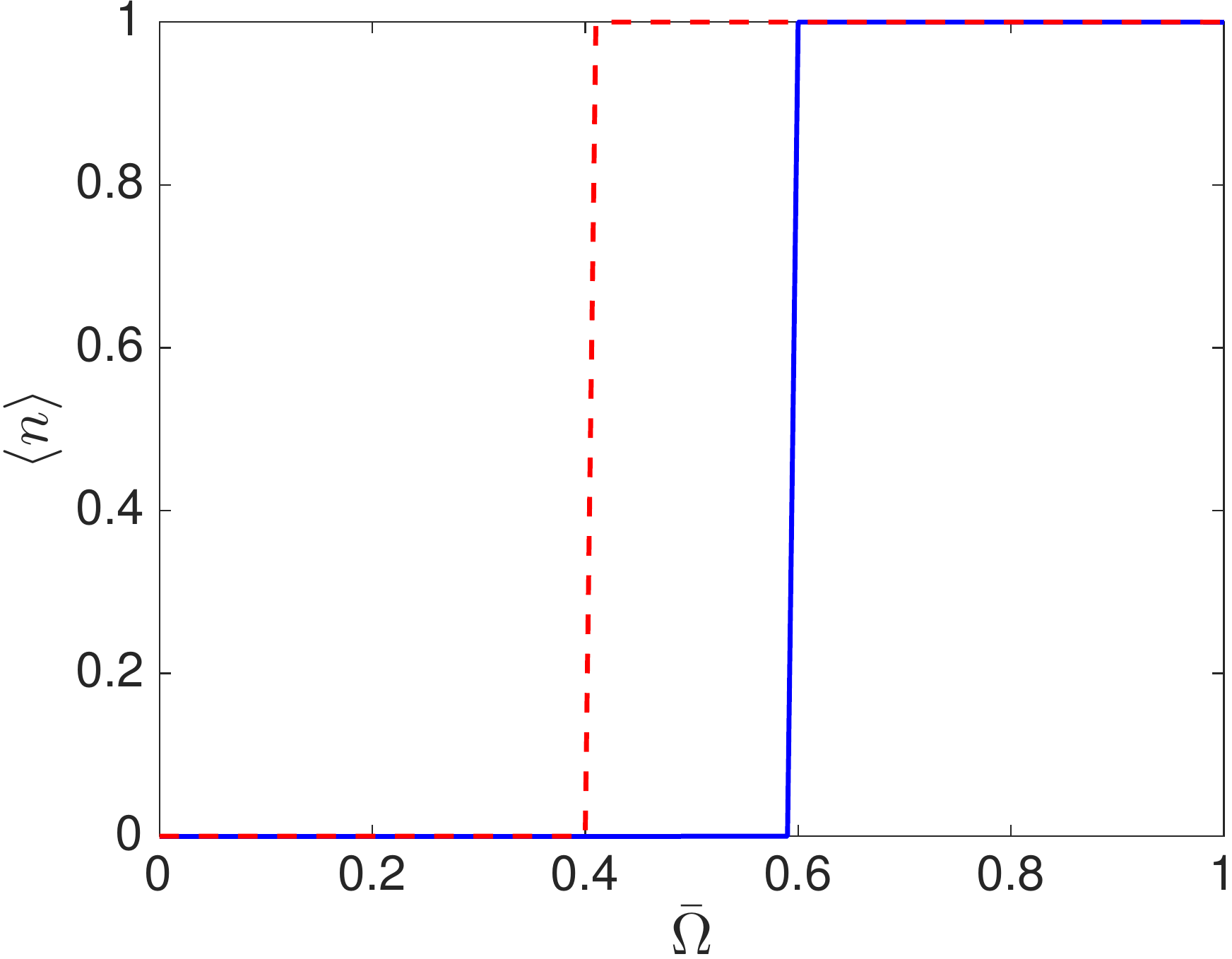}
  \caption[]{(Color online) The average momentum $\ev{n}$ via
    $\bar{\Omega}$ with parameters $N=1000$, $\bar{u}=0$, and
    $\bar{g}=0.0001$.~\label{fig:hyst0}}
\end{figure}

When our system is initially prepared in the ground state $\ket{N,0}$
by choosing $\bar{\Omega}=0$, the average angular momentum
$\ev{\hat{n}}=0$. As $\bar{\Omega}$ is quenched to the region $(0,\frac{1}{2})$,
$\ket{N,0}$ is still the ground state with $\ev{\hat{n}}=0$. As
$\bar{\Omega}$ is quenched to the region $(\frac{1}{2},\Omega_{c}^{+})$,
$\ket{N,0}$ is not a ground state but it is metastable, and we still
have $\ev{\hat{n}}=0$. As $\bar{\Omega}$ is quenched to the region
$(\Omega_{c}^{+},1)$, then $\ket{N,0}$ is not stable or metastable,
and the system evolves into the ground state $\ket{0,N}$ with
$\ev{\hat{n}}=1$. Similarly, we can analyze the case when
$\bar{\Omega}$ is quenched back from $1$ to $0$, and we find that
$\ev{\hat{n}}=1$ when $\bar{\Omega}\in(\Omega_{c}^{-},1)$, and
$\ev{\hat{n}}=0$ when $\bar{\Omega}\in(0,\Omega_{c}^{-})$. Therefore
it forms the hysteresis in the $\bar{\Omega}-\ev{\hat{n}}$ plane, see
Fig.~\ref{fig:hyst0} for a demonstration with $N=1000$ and
$\bar{g}=0.0001$.


\subsection{The case of $\bar{g}\protect\neq0,\bar{u}\protect\neq0$}
\label{sec:3d}

In this general case, we can not solve the eigen problem of
$\bar{H}_{S}$ analytically, and must resort to the numerical method.

\begin{figure}[htbp]
  \centering %
  \subfloat[][]{\label{fig:metas-a} \includegraphics[width=4cm]{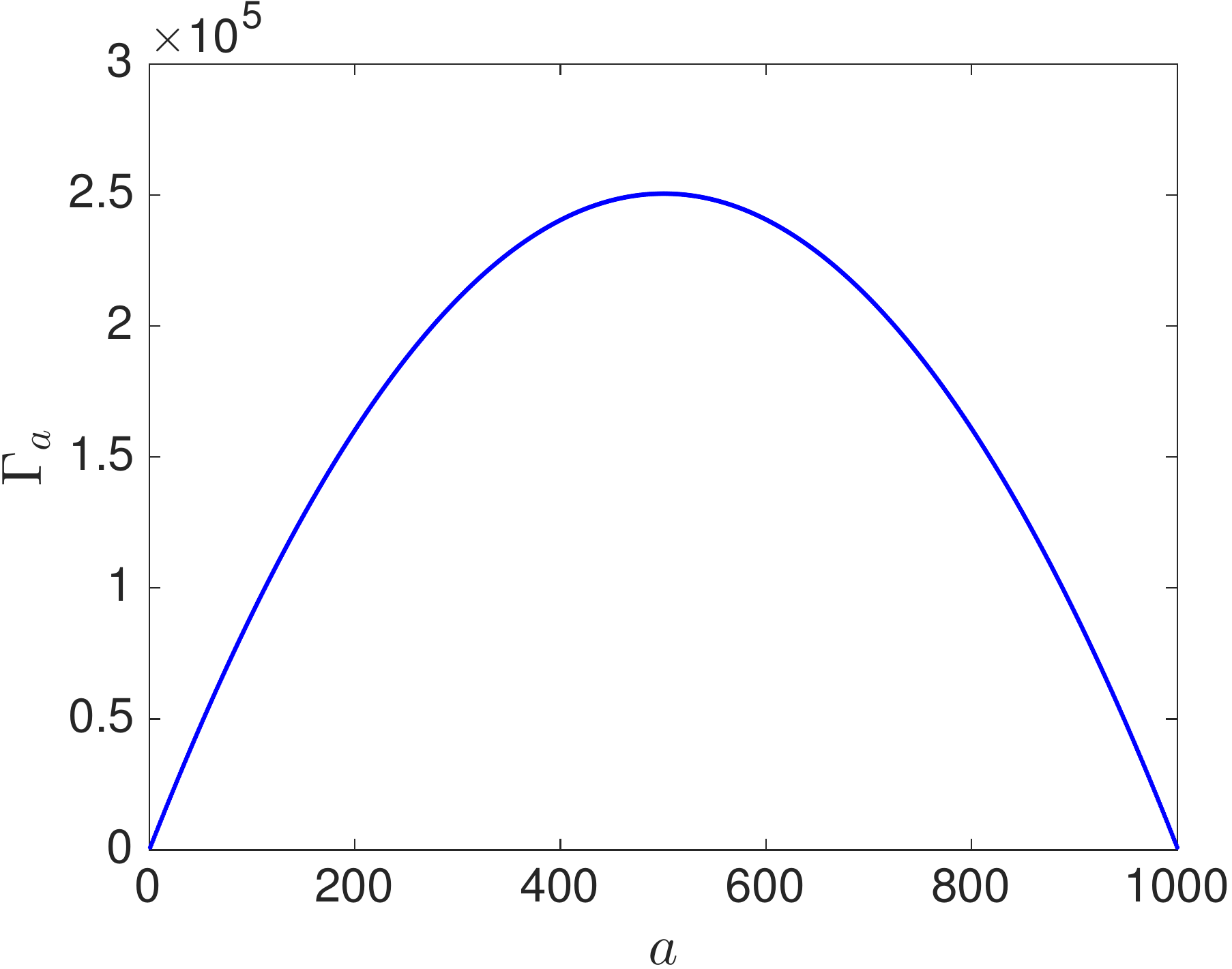}}
  \quad
  \subfloat[][]{\label{fig:metas-b} \includegraphics[width=4cm]{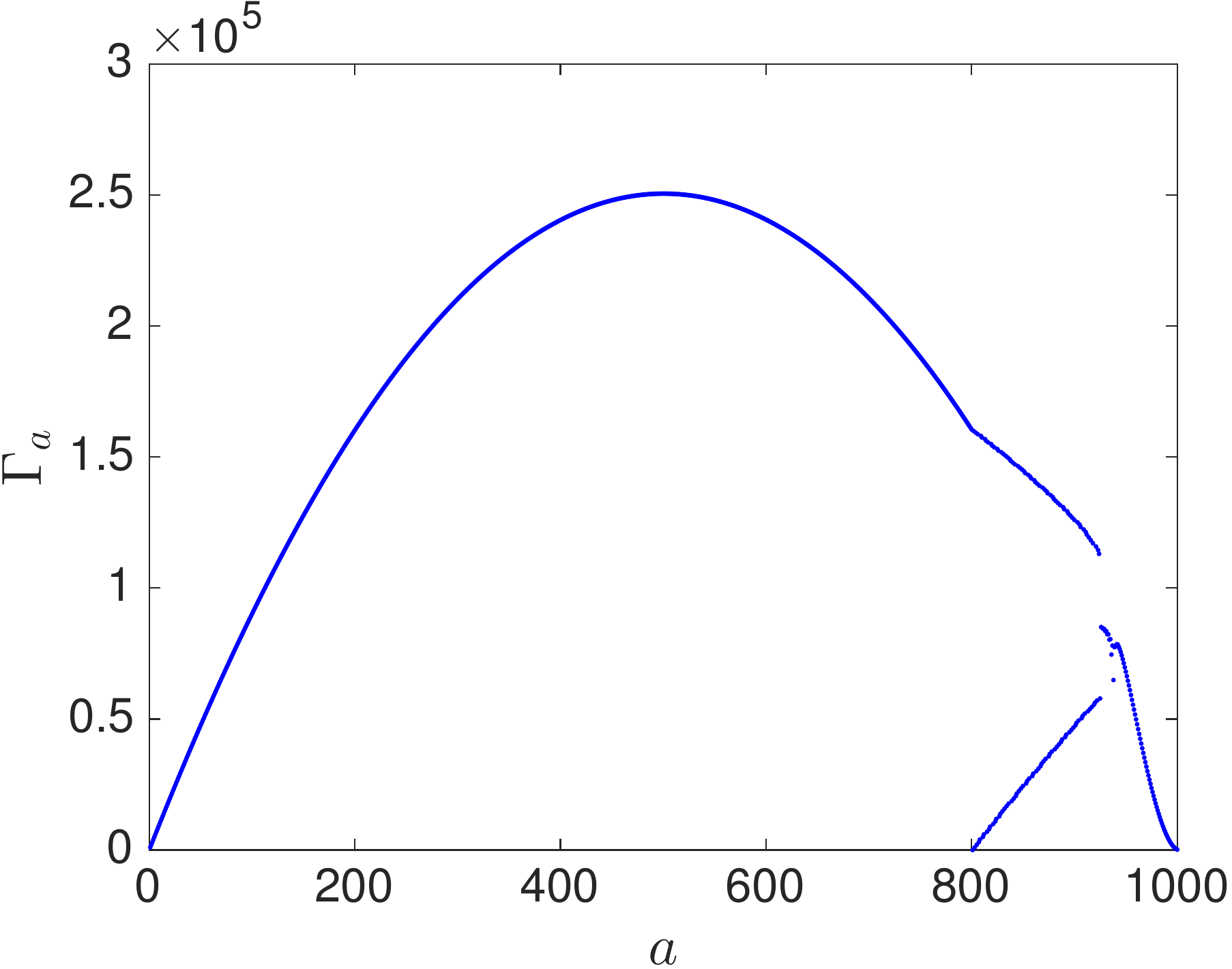}}
  \caption[The decay rates]{(Color online) The decay rates of excited
    states with parameters $N=1000$, $\bar{g}=0.0001$,
    $\bar{u}=0.0002$, and \subref{fig:metas-a} $\bar{\Omega}=0.35$;
    \subref{fig:metas-b} $\bar{\Omega}=0.42$.}~\label{fig:metas}
\end{figure}

First, we need to calculate all the eigenenergies and eigenstates of
$\bar{H}_{s}$, and determine whether there is a metastable state in
our system, which can be obtained from the decay rates of excited
states of $\bar{H}_{S}$. If there is a metastable state, then there is
an excited state with near zero decay rate. Here we give an example on
how to numerically confirm the existence of metastable states. When
$N=1000$, $\bar{g}=0.0001$, and $\bar{u}=0.0002$, the decay rates of
excited states with $\bar{\Omega}=0.35$ and $\bar{\Omega}=0.42$ are
shown in Fig.~\ref{fig:metas-a} and Fig.~\ref{fig:metas-b}
respectively. In Fig.~\ref{fig:metas-a} the two smallest decay rates
are $\Gamma_{1}=999.997949$ and $\Gamma_{1000}=999.872868$, which
implies that there does not exist metastable states in this case. In
Fig.~\ref{fig:metas-b}, the four smallest decay rates are given by
$\Gamma_{1}=999.96427$, $\Gamma_{803}=1000.597295$,
$\Gamma_{1000}=157.236302$, and $\Gamma_{801}\simeq 0$, which implies
that the $801$-th eigenstate is metastable.

\begin{figure}[htbp]
  \centering %
  \subfloat[][]{\label{fig:hyst1-a} \includegraphics[width=4cm]{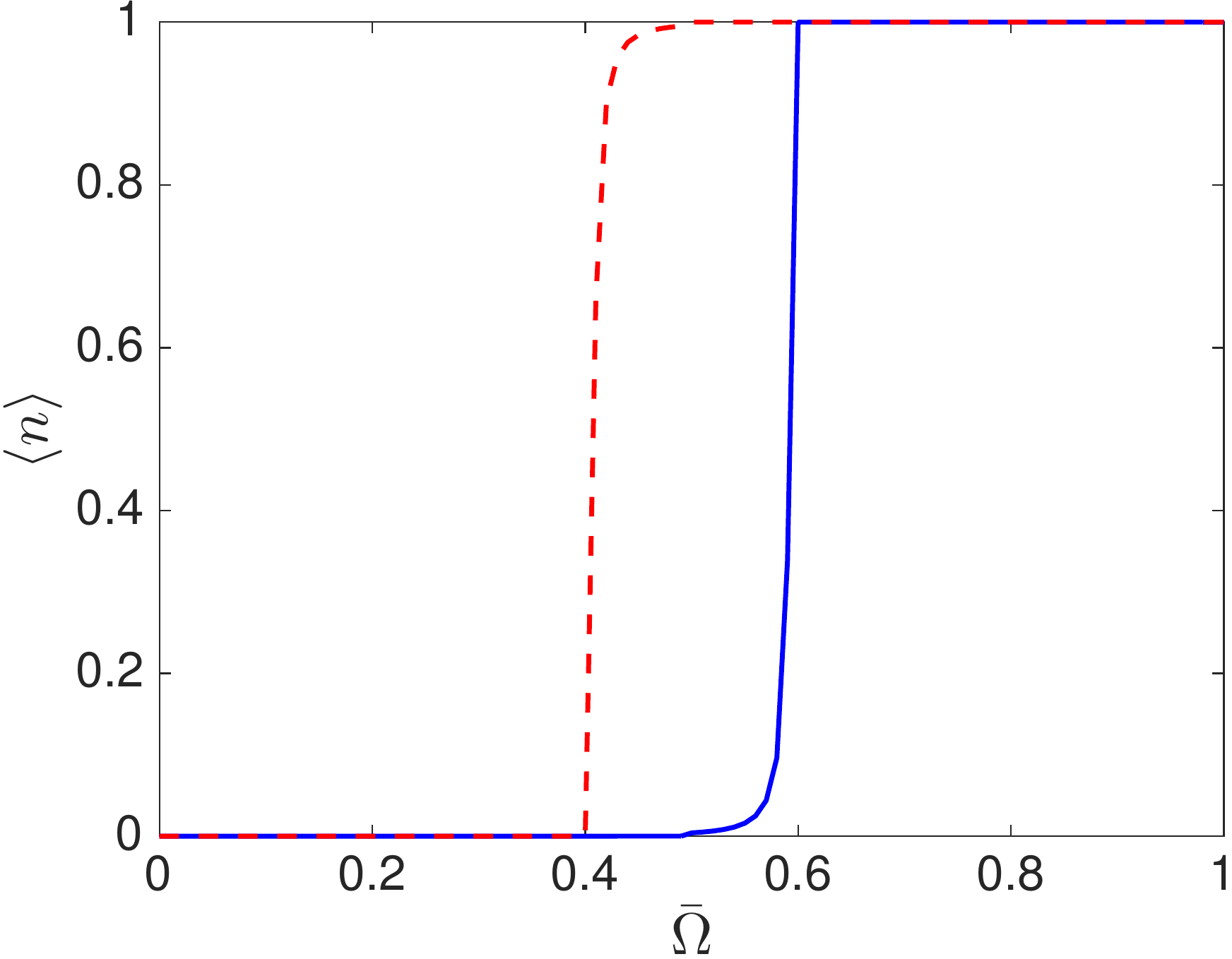}}
  \quad
  \subfloat[][]{\label{fig:hyst1-b} \includegraphics[width=4cm]{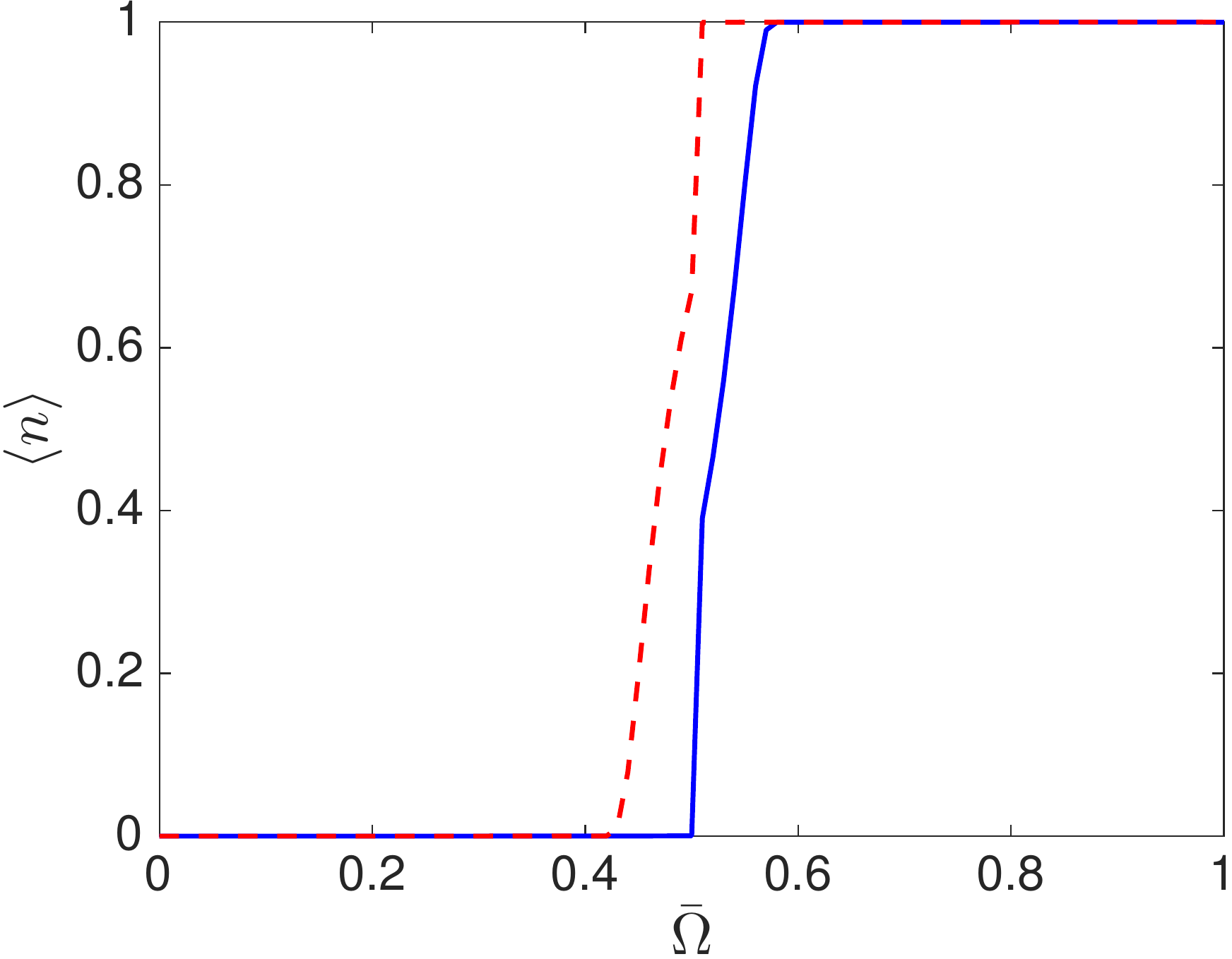}} \\
  \subfloat[][]{\label{fig:hyst1-c} \includegraphics[width=4cm]{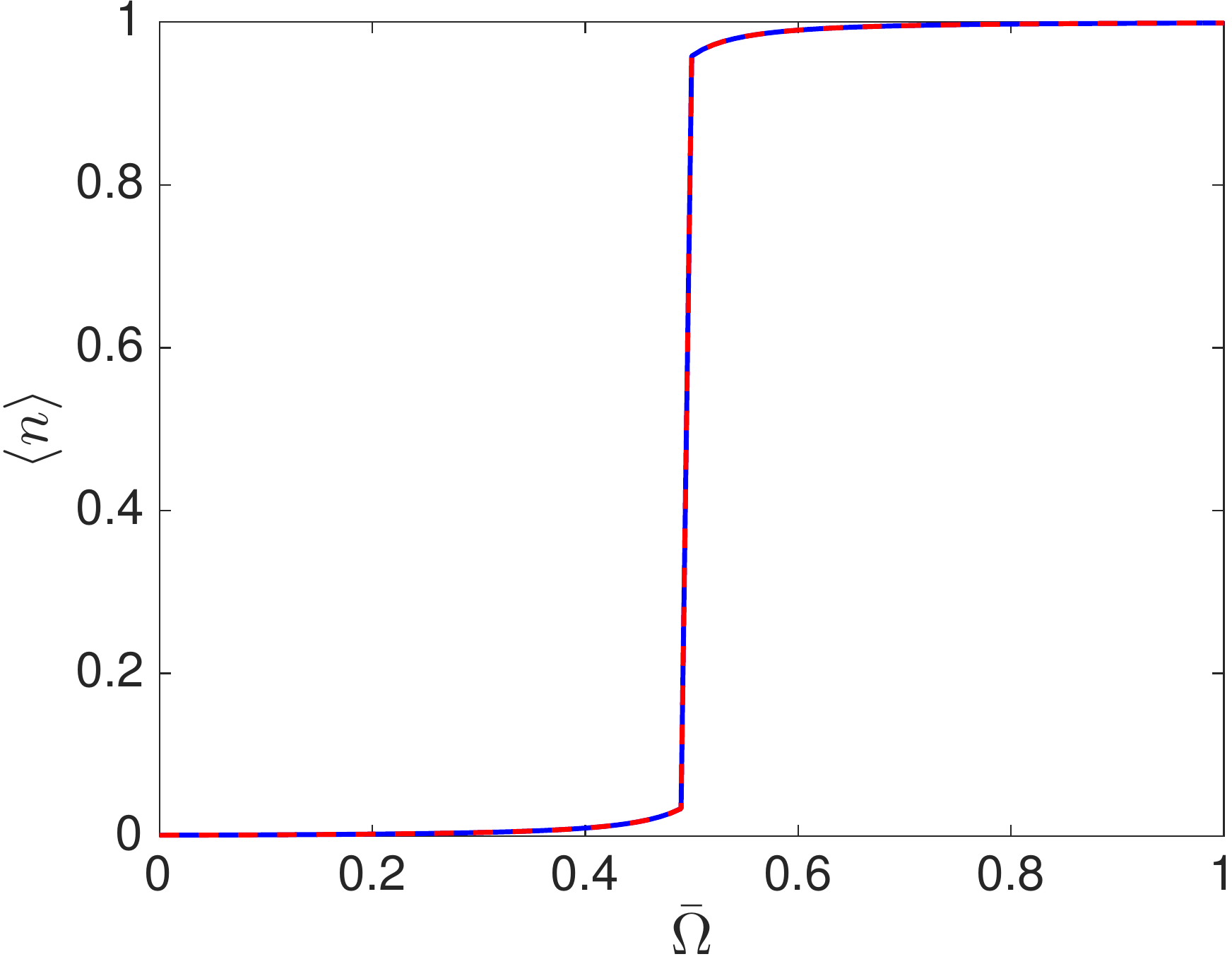}}
  \quad
  \subfloat[][]{\label{fig:hyst1-d} \includegraphics[width=4cm]{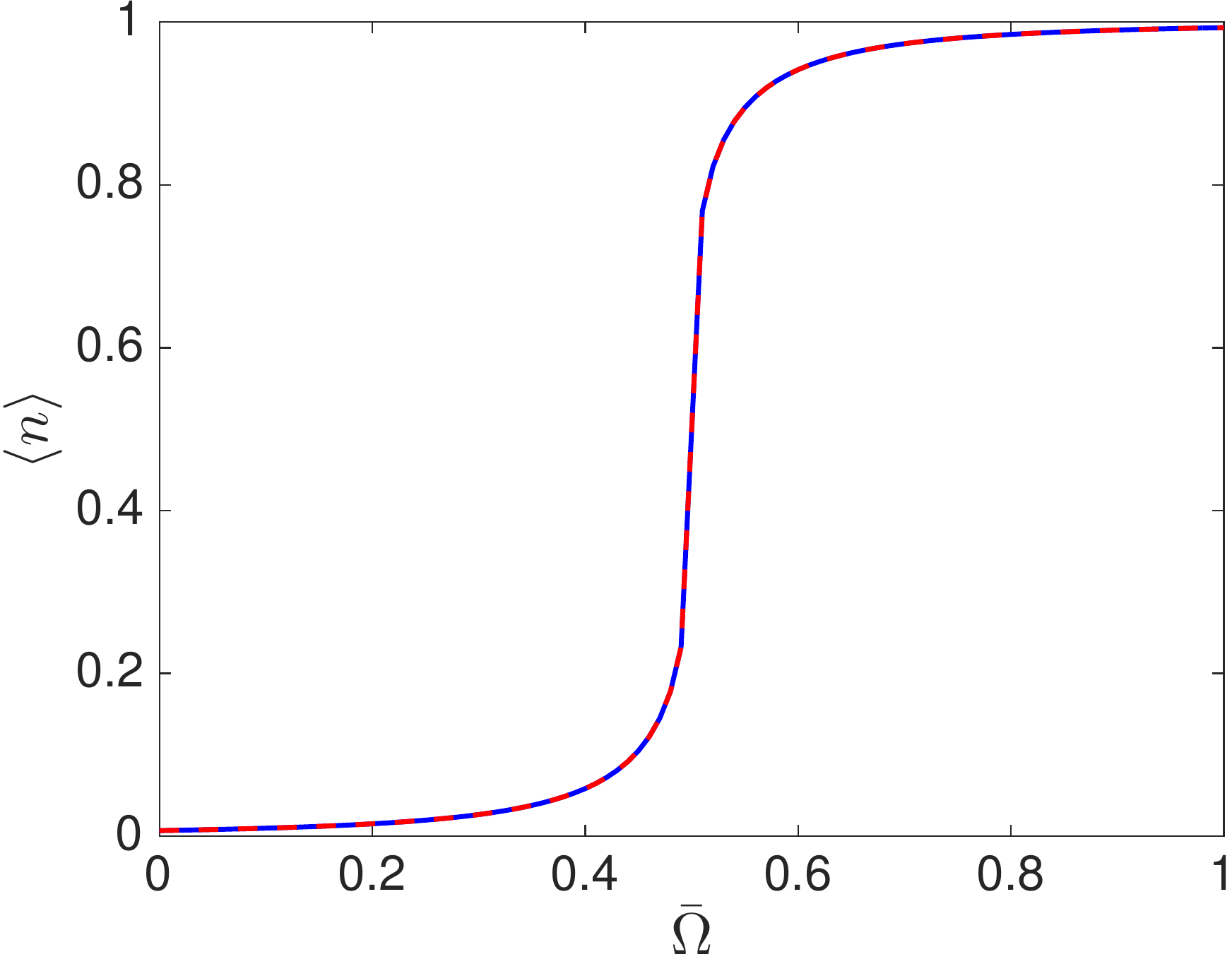}}
  \caption{(Color online) The average momentum $\ev{n}$ via
    $\bar{\Omega}$ with parameters $N=1000$, $\bar{g}=0.0001$, and (a)
    $\bar{u}=0.0002$; (b) $\bar{u}=0.002$; (c) $\bar{u}=0.02$; (d)
    $\bar{u}=0.05$.}~\label{fig:hyst1}
\end{figure}

Then we calculate the average momentum $\ev{n}$ as a function of
$\bar{\Omega}$ to show the hysterisis in our system. As shown in
Sec.~\ref{sec:3b}, the hysterisis will appear when $\bar{g}\neq 0$ and
$\bar{u}=0$ demonstrated in Fig.~\ref{fig:hyst0}. Now we examine how
the hysterisis changes with the increasing of $\bar{u}$ for given
parameters $N$ and $\bar{g}$, which is demonstrated in
Fig.~\ref{fig:hyst1} with different $\bar{u}$. In the case of $N=1000$
and $\bar{g}=0.0001$, there is a regular hysteresis in the plane of
$\bar{\Omega}-\ev{n}$ when $\bar{u}=0.0002$ shown in
Fig.~\ref{fig:hyst1-a}, which is smoother than that in
Fig.~\ref{fig:hyst0}. When $\bar{u}=0.002$, the hysterisis becomes
narrower demonstrated in Fig.~\ref{fig:hyst1-b}. When $\bar{u}=0.02$
in Fig.~\ref{fig:hyst1-c}, the hysterisis almost disappears. When
$\bar{\Omega}=0.05$ shown in Fig.~\ref{fig:hyst1-d}, the behavior of
$\ev{n}$ becomes similar as that in Fig.~\ref{fig:g0}, and the
hysteresis completely disappears.

As shown above, the appearance of metastable state is essential to the
formation of hysterisis. Thus it is natural to characterize the degree
of hysterisis in our system with the width of the region with
metastable states. Denote the region of $\bar{\Omega}$ with metastable
states with $[\Omega_{c}^{-},\Omega_{c}^{+}]$, and the degree of
hysterisis becomes
\begin{equation}
  \Delta\bar{\Omega}=\Omega_{c}^{+}-\Omega_{c}^{-}.
\end{equation}

\begin{figure}[htbp]
  \centering
  \includegraphics[width=6cm]{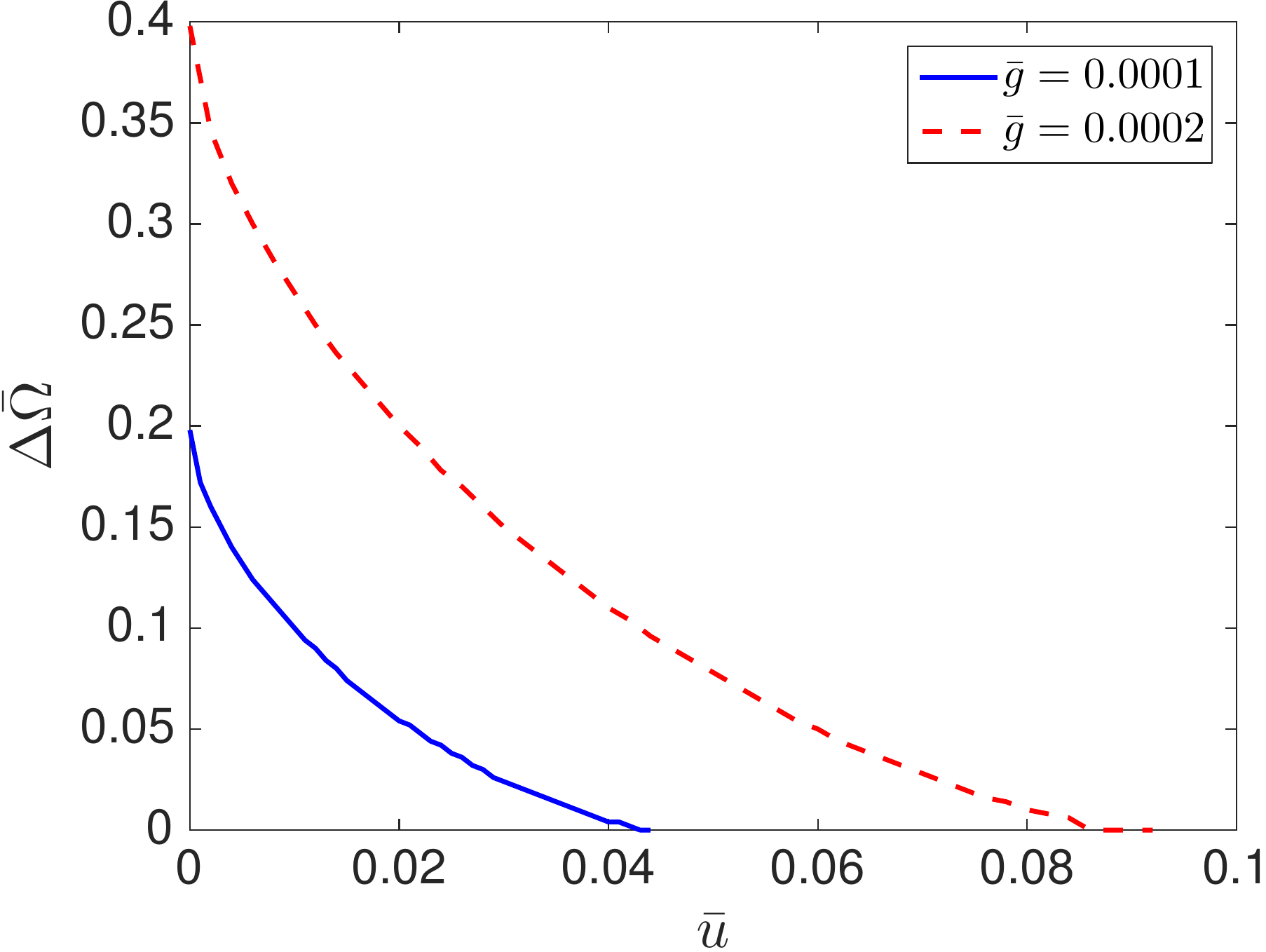}
  \caption{(Color online) The degree of hysterisis via $\bar{u}$ with
    parameter $N=1000$. The cases of $\bar{g}=0.0001$ and
    $\bar{g}=0.0002$ are denoted by the blue solid line and the red
    dashed line respectively.}~\label{fig:hystdeg}
\end{figure}

Based on the concept of the degree of hysteresis, we investigate how
$\bar{u}$ changes the degree of hysteresis for given $N$ and
$\bar{g}$, which is given in Fig.~\ref{fig:hystdeg}. In particular, it
defines a critical point $\bar{u}_{c}$ such that $\bar{u}<\bar{u}_{c}$
is a condition for the appearance of hysteresis. For example, when
$N=1000$ and $\bar{g}=0.0001$, the critical point $\bar{u}_{c}=0.046$.

\begin{figure}[htbp]
  \centering
  \includegraphics[width=8cm]{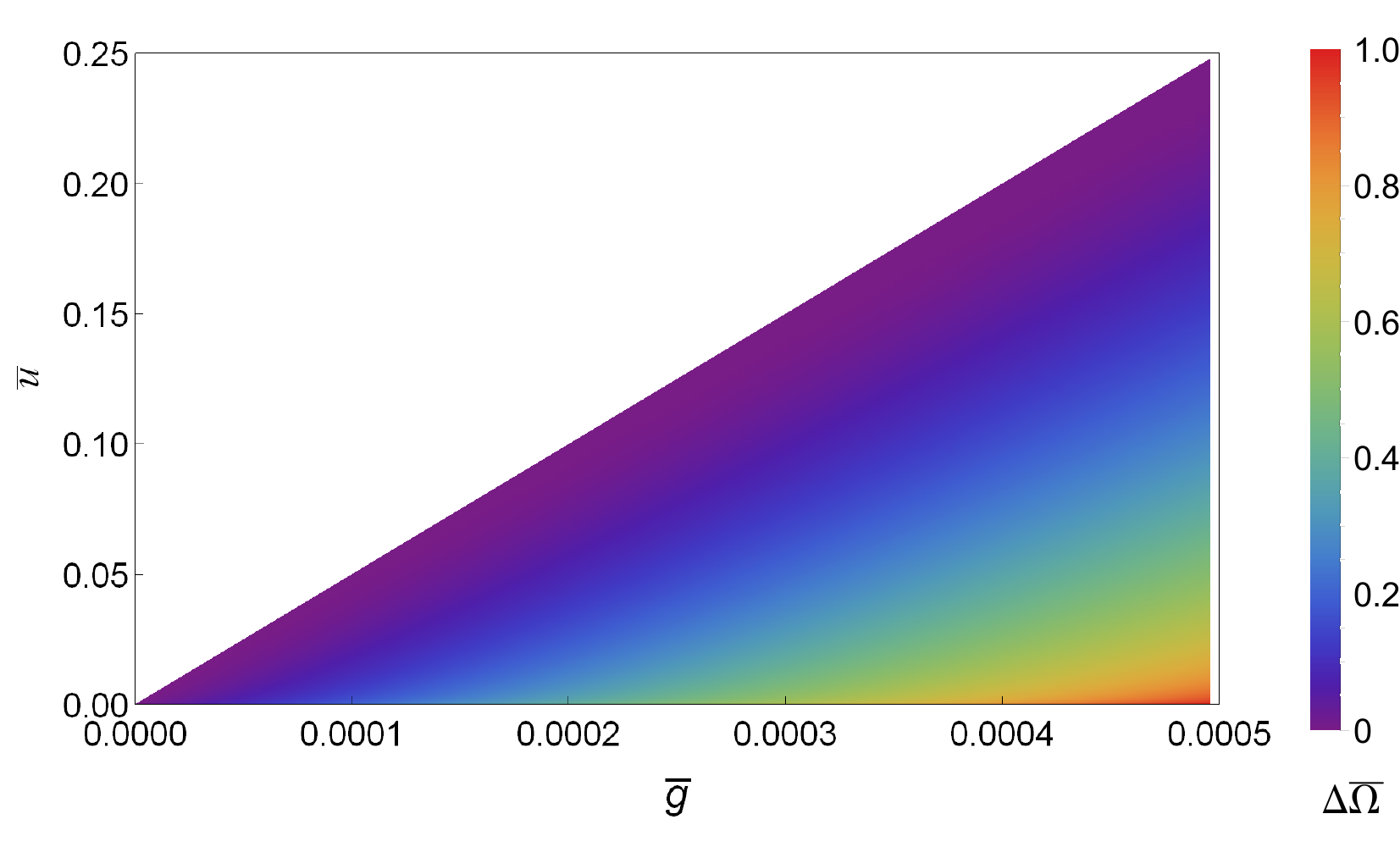}
  \caption{(Color online) The parameter regions where
    $\Delta\bar{\Omega}>0$ and the degree of hysteresis for different
    parameters with $N=1000$. }~\label{fig:hystgu}
\end{figure}

Furthermore, we present the degree of hysteresis $\Delta\bar{\Omega}$
for different parameters when $N=1000$. Fig.~\ref{fig:hystgu} suggests
that $\Delta\bar{\Omega}$ is determined by the competition between
$\bar{g}$ and $\bar{u}$: the former tends to increase
$\Delta\bar{\Omega}$ while the latter decrease it. Also, we notice
that $\bar{u}_{c}(\bar{g})\propto \bar{g}$( the boundary between the
parameter region where $\Delta\bar{\Omega}>0$ and the blank space in
Fig.~\ref{fig:hystgu}). In fact, for $\forall \bar{g}$, if $\bar{u}$
is set to be $\bar{u}_{c}(\bar{g})$,
$\Omega_{c}^{+}=\Omega_{c}^{-}=1/2$, it means that
$\bar{u}_{c}(\bar{g})$ is determined by the energy band structure of
Hamiltonian (\ref{hab}) for $\bar{\Omega}=1/2$,
i.e.,$\bar{H}=N/8+\bar{g}a_{0}^{\dagger}a_{0}a_{1}^{\dagger}a_{1}+\bar{u}(a_{0}^{\dagger}a_{1}+a_{1}^{\dagger}a_{0})$,
it's obvious that the energy band structure is determined by
$\bar{u}/\bar{g}$, leading to $\bar{u}_{c}(\bar{g})\propto \bar{g}$.

\section{Discussion and summary}

In our present treatment of hysteresis, we focuses on the
instantaneous change of $\bar{\Omega}$, which simplifies our
calculation of hysteresis and simulates well with the present
experiment. However, it is worthy to point out that this is not the
unique process to observe the hysterisis in our system. For example,
another interesting process for this purpose is the adiabatic change
of $\bar{\Omega}$ with a slow variant rate. In this adiabatic
process, the many-body Landau-Zener tunneling~\cite{CZen,LDLan,CWit}
may be involved to affect the hysteresis, which needs to be
investigated further in future.

\vspace{0.5cm}

In summary, we have presented a microscopic theory of the hysteresis
in an atomic BEC\@, in which a two-mode model is used to describe the
BEC system in a ring with an external potential, and an ensemble of
bosons is introduced to act as the environment of the BEC system. We
find that such a simple two-mode model captures the essence of
hysteresis, an effect from the interactions between atoms and the
dissipation induced by its environment. In particular, the existence
of an metastable state is essential for the formation of hysteresis in
our system. More precisely, the steady state depends on the initial
state when a metastable state appears, which leads to the hysteresis
in the average momentum $\ev{\hat{n}}$ as a function of the rotation
frequency $\bar{\Omega}$. We also find that the parameter of the
external potential $\bar{u}$, which controls the tunneling between the
two modes we consider, weakens the effect of hysterisis in our system.
In particular, for a given particle number $N$ and the interaction
constant $\bar{g}$, there is a critical potential $\bar{u}_{c}$ such
that if the potential is beyond the critical value, the hysteresis
will be completely destroyed. We numerically give the critical
potential parameter $\bar{u}_{c}$ for different interaction parameter
$\bar{g}$ when $N=1000$, and observe that
$\bar{u}_{c}\propto \bar{g}$, which is also proved theoretically. It
is worthy to point out that the predictions of our microscopic theory
are consistent with the observations in the present experiment. We
hope that our theory of hysteresis will promote our understandings of
hysteresis at the microscopic level, and gives more rich and precise
predictions beyond the mean field theory.


\begin{acknowledgments}
  This work is supported by NSF of China (Grant Nos. 11475254 and
  11175247) and NKBRSF of China (Grant Nos. 2012CB922104 and
  2014CB921202).
\end{acknowledgments}

\end{document}